\documentstyle[aps,twocolumn,epsf,psfig]{revtex}

\begin{document}
\draft

\title {Exact weak-coupling radius of the Holstein polaron in one, two, and three dimensions}

\author{Aldo~H.~Romero${^{1,3}}$, David W. Brown${^2}$ and Katja Lindenberg${^3}$}

\address
{${^1}$
Department of Physics,\\
University of California, San Diego, La Jolla, CA 92093-0354}

\address
{${^2}$
Institute for Nonlinear Science,\\
University of California, San Diego, La Jolla, CA 92093-0402}

\address
{${^3}$
Department of Chemistry and Biochemistry,\\
University of California, San Diego, La Jolla, CA 92093-0340} 

\date{\today} 

\maketitle

\begin{abstract}

We apply weak-coupling perturbation theory to the Holstein molecular crystal model in order to compute an electron-phonon correlation function characterizing the shape and size of the polaron lattice distortion in one, two, and three dimensions.
This correlation function is computed exactly to leading order in the electron-phonon coupling constant, permitting a complete description of correlations in any dimension for both isotropic and arbitrarily anisotropic cases.
Using this exact result, the width of the polaron is characterized along arbitrary directions.
The width of the polaron thus determined disagrees in every dimension with some well-known characterizations of polarons, signalling in particular the breakdown of the adiabatic approximation and the characterizations of self-trapping associated with it.

\end{abstract}


\narrowtext

As quantum quasiparticles, polarons (of whatever particular variety) describe the states excitations correlated with the deformation or polarization quanta of a host medium.
Often characterized in terms such as "large" and "small" and "free" and "self-trapped", perhaps the single most intuitively-accessible polaron property is its size.
Our principal interest in this paper is the "polaron radius", which is, however, not a uniquely-definable quantity.
In many approximate approaches, one encounters polarons characterized by spatially-localized semi-classical wavefunctions for which a radius appears to be easily defined through the variance of the squared probability amplitude of a localized electron.
Such localized polarons are often described as "self-trapped".
When such localized wavefunctions are not found, or are not found to be energetically favored, the polaron is often said to be delocalized or "free" and the polaron radius is then said to be infinite \cite{Toyozawa61,Toyozawa63,Emin73,Sumi73,Emin76,Ueta86,Kabanov93,Toyozawa80a,Silinsh94,Song96,Toyozawa97}.

These characterizations of polaron radii do not generalize to energy band theory where, owing to the translational invariance of the crystal lattice, all states are delocalized.
When a Bloch state is constructed from localized functions, the "identity" of the localized function is lost and one can no longer characterize the properties of the quantum state in terms uniquely associated with the local function used to convey local correlation structure into the delocalized quantum state \cite{Brown95a}.
These local properties are probed with correlation functions that reveal the internal structure implicit in the delocalized state \cite{Alexandrov95}.

In this paper, we focus on one such correlation function that has been long and widely used to characterize polaron size in D dimensions:
\begin{equation}
C_{\vec{r}}^{[D]} = \langle \hat{C}_{\vec{r}}^{[D]} \rangle = \frac 1 {2g} \langle \sum_{\vec{n}} a_{\vec{n}}^{\dagger} a_{\vec{n}} (b_{\vec{n}+\vec{r}}^{\dagger} + b_{\vec{n}+\vec{r}} ) \rangle  ~.
\label{eq:corrfunc}
\end{equation}
This function can be viewed as measuring the shape of the polaron lattice distortion around the instantaneous position of the electron, or, essentially equivalently in view of the strongly local character of the electron-phonon coupling, as an image of the electron density one associates with a localized polaron in semi-classical approaches.
Indeed, a common finding in the adiabatic approximation is that this correlation function can be expressed as the real-space autocorrelation function of a localized electron density, whose spatial variance $\sigma^2$ is simply proportional to the square of the polaron radius $R$.
In 1-D, one finds the particularly common result
\begin{equation}
\sigma^2 \propto R^2 \propto \left( \frac {J} {g^2 \hbar \omega} \right) ^2
\label{eq:radius}
\end{equation}
that captures in one very simple relationship between the electron transfer integral $J$, the phonon frequency $\omega$, and the electron-phonon coupling constant $g$ much of what is commonly considered characteristic of polarons in 1-D.

We assert that the 1-D characterization of polaron size as contained in (\ref{eq:radius}) and the 2-D and 3-D characterization of polarons as distinguishably "free" or "self-trapped" fail to capture the size dependence of polarons in the weak coupling regime.
We apply weak-coupling perturbation theory to derive the exact form of the correlation function (\ref{eq:corrfunc}) in 1-D, 2-D, and 3-D, and from this we derive the exact real-space variance that characterizes the polaron radius in any direction.
This result is incompatible with adiabatic theory in 1-D, 2-D, and 3-D, but is compatible with results of other approaches not limited by the adiabatic approximation.
This incompatibility ultimately holds significant implications for the notions of polaron size and the concept of self-trapping generally.

We use the Holstein Hamiltonian \cite{Holstein59a,Holstein59b}
\begin{equation}
\hat{H} = \hat{H}_{kin} + \hat{H}_{ph} + \hat{H}_{int} ~,
\end{equation}
\newpage
\begin{eqnarray}
\hat{H}_{kin} &=& - \sum_{ \vec{n} }
\left[
J_a a_{\vec{n}}^{\dagger}
( a_{\vec{n}+\vec{\epsilon}_a} + a_{\vec{n}-\vec{\epsilon}_a} ) \right. \nonumber \\
&&
~~~~~ ~ + J_b a_{\vec{n}}^{\dagger}
( a_{\vec{n}+\vec{\epsilon}_b} + a_{\vec{n}-\vec{\epsilon}_b} ) \nonumber \\
&&
\left.
~~~~~ ~ + J_c a_{\vec{n}}^{\dagger}
( a_{\vec{n}+\vec{\epsilon}_c} + a_{\vec{n}-\vec{\epsilon}_c} ) \right]
 ~, \\
\hat{H}_{ph} &=& \hbar \omega \sum_{\vec{n}} b_{\vec{n}}^{\dagger} b_{\vec{n}} ~, \\
\hat{H}_{int} &=& - g \hbar \omega \sum_{\vec{n}} a_{\vec{n}}^{\dagger}
a_{\vec{n}} ( b_{\vec{n}}^{\dagger} + b_{\vec{n}} ) ~,
\end{eqnarray}
in which $a_{\vec{n}}^\dagger$ creates a single electronic excitation in the rigid-lattice Wannier state at site ${\vec{n}}$, and $b_{\vec{n}}^\dagger$ creates a quantum of vibrational energy  in the Einstein oscillator at site ${\vec{n}}$.
All sums are understood to run over the entire $D$-dimensional lattice of edge-length $N$.
Because there is no phonon dispersion in this model, and because the electron-phonon coupling is strictly local, $\hat{H}_{ph}$ and $\hat{H}_{int}$ are directly affected by the dimensionality and structure of the lattice only through the number of terms contained in each sum.
It is in $\hat{H}_{kin}$ where lattice dimensionality and structure have their greatest influence; the $J_i$ are the nearest-neighbor electronic transfer integrals along the primitive crystal axes, and the $\hat{\epsilon}_i$ are unit vectors associated with the primitive translations.
We emphasize that the discrete, integer $D$-tuples $\vec{n}$ label primitive unit cells along the primitive axes $a$, $b$, and $c$, such that the discrete vector analysis involved in this paper is formally independent of lattice structure.
As such, the above model encompasses all Bravais lattices, with the different lattice structures appearing only in the relative values of the three hopping integrals $J_a$, $J_b$, and $J_c$.

When considering less than three dimensions, one may be legitimately concerned with possible distinctions between a pure 2-D scenario, for example, and cases of "mixed" dimensionality wherein the electronic degrees of freedom may be essentially 2-D in layers of a 3-D medium.
Within the scope of the results here presented there is no distinction between "pure" and "mixed" dimensionalities so that the {\it electronic} dimensionality is the only relevant dimensionality.
The 2-D case is therefore obtained by setting any one $J_i$ equal to zero, and the 1-D case is obtained by setting any two $J_i$ equal to zero.

The core issue of concern in this paper is evident in a small but powerful way in the expected value of the interaction energy in the polaron ground state ($\vec{\kappa}=0$), $E_{int} = \langle 0 | \hat{H}_{int} | 0 \rangle$.
This energy is related to the center value of the correlation function (\ref{eq:corrfunc}) through
\begin{equation}
E_{int} = -2 g^2 \hbar \omega {C}_{0}^{[D]} = -2 \epsilon_p {C}_{0}^{[D]} ~,
\end{equation}
in which $\epsilon_p$ is the small polaron binding energy in the extreme strong-coupling limit.
Any non-zero value of $E_{int}$ implies a non-zero value of ${C}_0^{[D]}$, and, since $E_{int}$ is non-vanishing at any finite $J_i$ and $g$, it necessarily follows that ${C}_0^{[D]} $ is non-zero in any interesting case.
However, since the correlation function (\ref{eq:corrfunc}) satisfies the sum rule $\sum_{\vec{r}} {C}_{\vec{r}}^{[D]} = 1$ for all $J_i$ and $g$ in any dimension, this non-trivially finite value of ${C}_0^{[D]} $ implies the generic existence of non-trivial local character in the electron-phonon correlations that define polaron structure.
This is concluded without appeal to any approximation method or specific calculation; it is a general consequence of a non-zero electron-phonon interaction energy in the polaron model.
This limited, but broad conclusion already appears to contradict some expectations for polaron structure in two and three dimension where infinite-radius "free" states are suggested to prevail in the weak-coupling regime.
It is possible, in principle, for localized correlations to be compatible with an infinite polaron radius if the localization is sufficiently weak that its moments are all infinite.
To fully settle this question, direct computation is required.

\begin{figure*}[htb]
\begin{center}
\leavevmode
\epsfxsize = 3.5in
\epsffile{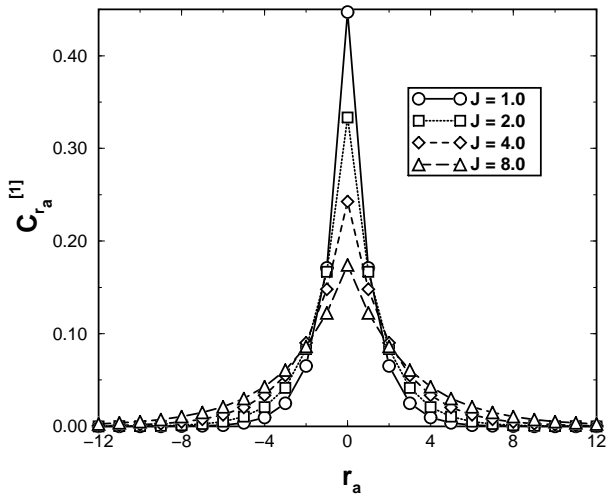}
\end{center}
\caption
{Weak-coupling 1-D correlation function ${C}_{r_a}^{[1]}$ for $J_a / \hbar \omega = 1$, $2$, $4$, and $8$.}
\label{fig:c1dwcpt}
\end{figure*}

We now pursue Rayleigh-Schr\"{o}dinger perturbation theory \cite{Alexandrov95,Nakajima80,Mahan93} in the weak-coupling regime identifying the unperturbed Hamiltonian as $\hat{H}_0 = \hat{H}_{kin} + \hat{H}_{ph}$ and the perturbation as $\hat{H}' = \hat{H}_{int}$
\begin{equation}
| \psi [D] \rangle = | \psi ^{(0)} [D] \rangle + | \psi^{(1)} [D] \rangle + ... ~.
\end{equation}
General $\vec{\kappa}$-dependences can be obtained, but for our purposes we require only the $\vec{\kappa} = 0$ results
\begin{eqnarray}
| \psi^{(0)} [D] \rangle &=& a_{\vec{k}=0}^{\dagger} | 0 \rangle ~, \\
| \psi^{(1)} [D] \rangle &=&
- \frac {g \hbar \omega} {\sqrt{N^D}}
\sum_{\vec{q}} \frac
{a_{\vec{k} = -\vec{q}}^{\dagger} b_{\vec{q}}^{\dagger} | 0 \rangle }
{\{E(0) - [E(-\vec{q} ) + \hbar \omega ] \}} ~,
\end{eqnarray}
where
\begin{equation}
 E(\vec{k}) = -2 \sum_{i=1}^D J_i \cos k_i
\end{equation}
is the free electron band energy at finite $\vec{k}$.
Consequently, one finds that
\begin{equation}
{C}_{\vec{r}}^{[D]} = 2 \langle \psi ^{(0)} [D] \, | ~ \hat{C}_{\vec{r}}^{[D]} ~ | \, \psi ^{(1)} [D] \rangle + O\{ g^2 \} ~,
\end{equation}
which to leading order in $g$ (equivalent to order $g^2$ in the ground state energy) yields
\begin{equation}
{C}_{\vec{r}}^{[D]} = - \frac {\hbar \omega} {N^D} \sum_{\vec{q}} \frac {e^{i \vec{q} \cdot \vec{r} }} {\{E(0) - [E(- \vec{q}) + \hbar \omega] \} } ~,
\end{equation}
which can be further developed into the appealing expressions
(with obvious generalization to $D > 3$)
\begin{eqnarray}
{C}_{\vec{r}}^{[1]} &=& \hbar \omega \int_0^{\infty} \!\! du ~ e^{-(2 J_a + \hbar \omega )u} I_{r_a} (2J_a u) ~, \label{eq:c1d} \\
{C}_{\vec{r}}^{[2]} &=& \hbar \omega \int_0^{\infty} \!\! du ~ e^{-(2J_a + 2J_b + \hbar \omega )u} \nonumber \\
&& ~~~~~ ~~~~~ \times I_{r_a} (2J_a u) I_{r_b} (2J_b u) ~, \label{eq:c2d} \\
{C}_{\vec{r}}^{[3]} &=& \hbar \omega \int_0^{\infty} \!\! du ~ e^{-(2J_a + 2J_b + 2J_c + \hbar \omega )u} \nonumber \\
&& ~~~~~ ~~~~~ \times I_{r_a} (2J_a u) I_{r_b} (2J_b u) I_{r_c} (2J_c u)  ~.
\label{eq:c3d}
\end{eqnarray}
In which $I_n ( z )$ is the modified Bessel function of the first kind.
Note that when any one $J_i$ is set to zero, the contribution from the $i$ axis is reduced to a delta function $\delta_{r_i 0}$, showing one way in which the 1-D and 2-D cases can be recovered straightforwardly from the 3-D case.
It is also the case, however, that for any set of $J_i$, summing $ {C}_{\vec{r}}^{[3]} $ over one $r_i$ recovers ${C}_{\vec{r}}^{[2]} $, and further summing the latter over either of the remaining $r_i$ recovers ${C}_{\vec{r}}^{[1]} $; thus, the correlation function that results from averaging out the transverse dependences of a multi-dimensional correlation function results in the same 1-D correlation function that would have resulted from setting the transverse $J_i$ equal to zero.
Figure~\ref{fig:c1dwcpt} shows several examples of the 1-D result (\ref{eq:c1d}).

\begin{figure*}[htb]
\begin{center}
\leavevmode
\epsfxsize = 3.5in
\epsffile{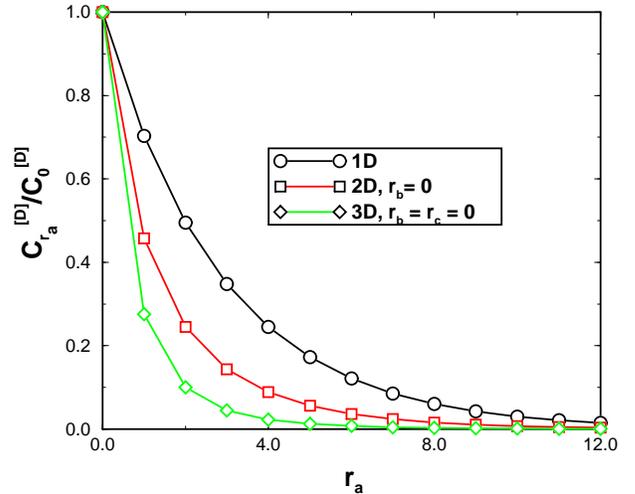}
\end{center}
\caption
{Correlation decay along the $a$ axis in the weak-coupling limit in 1-D, 2-D, and 3-D as measured by ${C}_{r_a \hat{\epsilon_a}}^{[D]} / {C}_0^{[D]} $ for $J_a = J_b = J_c = 8 \hbar \omega $.}
\label{fig:c123dwcpt}
\end{figure*}

If instead of averaging out transverse dimensions one makes a 1-D cut through a selected portion of the correlation function (e.g. $\vec{r}_b = \vec{r}_c = 0$), one finds something other than 1-D character.
Figure~\ref{fig:c123dwcpt} illustrates the dimensionality dependence of the correlation function by comparing 1-D transections of the correlation function in 1-D, 2-D, and 3-D for an isotropic case.
The 1-D cut through the D-dimensional correlation function is normalized by the value of that correlation function at $\vec{r} =0$, so that what is displayed is a fair relative measure of how rapidly correlations decay with radial distance from the center of the polaron.
As might be anticipated from the form of (\ref{eq:c1d}) - (\ref{eq:c3d}), correlations are seen to decay more rapidly with increasing real-space dimensionality.

Characterizing this multi-dimensional correlation function in terms of a width measure involves selecting a direction along which to contract a variance tensor;
thus, given an appropriate unit vector $\vec{\mu}$, one writes
\begin{equation}
\sigma_{\vec{\mu}}^2 = \vec{\mu} \cdot \mbox{ \boldmath $\sigma^2$ } \cdot \vec{\mu} ~.
\end{equation}

\noindent
The components of $\mbox{ \boldmath $\sigma^2$ }$ can be computed from
\begin{equation}
\sigma^2_{ij} = \sum_{\vec{r}} r_i r_j {C}_{\vec{r}}^{[D]} ~,
\end{equation}
where $i,j = a$, $b$, or $c$, resulting in
\begin{equation}
\mbox{\boldmath $\sigma^2$ } = \left( 
\begin{array}{ccc}
\sigma_{aa}^2 & 0 & 0 \\
0 & \sigma_{bb}^2 & 0 \\
0 & 0 & \sigma_{cc}^2
\end{array}
\right) ~,
\end{equation}
where
\begin{eqnarray}
\sigma_{ii}^2 &=& \hbar \omega \int_0^{\infty} \!\! du ~ e^{- \hbar \omega u} ~ \sum_{r_i} ~ r_i^2 \, e^{- 2J_i u} I_{r_i} (2J_i u) \\
&=& \frac {2J_i} {\hbar \omega} ~.
\end{eqnarray}
Thus, along each of the primitive crystallographic axes, the real-space variance is simply proportional to the electron transfer integral along that axis, and in a general direction is just the appropriate mixture determined by rotation.

Contrary to much prevailing opinion, this result shows that in the weak-coupling regime:
i) there are no significant qualitative or quantitative differences between 1-D, 2-D, and 3-D polaron radii,
ii) the polaron radius in 2-D and 3-D is {\it not} infinite, and
iii) the polaron radius does not scale as $J/g^2 \hbar \omega$ as commonly expected, but as $\sqrt{J/ \hbar \omega}$.

Since these are exact results of weak-coupling perturbation theory, the only way for these conclusions to be invalidated would be for perturbation theory to fail {\it dramatically} at infinitesimal $g$; however, there is no evidence of such potential and catastrophic failure to be found.
Weak-coupling perturbation theory is well-behaved in the vicinity of the Brillouin zone center where both the unperturbed and perturbed states are well-separated from the one-phonon continuum ($| \vec{\kappa} | < \kappa_c$, where $2J_{max}[ 1- \cos \kappa_c ] = \hbar \omega$).
In 1-D, the results of weak-coupling perturbation theory are broadly validated by a variety of high-quality, non-perturbative techniques demonstrating the qualitative and quantitative accuracy of the former to non-trivially finite coupling strengths \cite{DeRaedt83,DeRaedt84,Lagendijk85,Alexandrov94a,Marsiglio95,Zhao97a,Zhao97b,Brown97b,Wellein97a,Jeckelmann98a,Romero98a}.
Results are more scarce in 2-D and 3-D; however, where reliable non-perturbative results are available in the appropriate regime, they support rather than contradict the results of weak-coupling perturbation theory.

On the other hand, methods invoking the adiabatic approximation do not enjoy such extensive corroboration in the weak coupling regime.
This has been explained through a demonstration that one of the key features of the adiabatic approximation -- a locking relation between electron and phonon coordinates -- eventually dematerializes as electron-phonon coupling is decreased at finite $J/\hbar \omega$ \cite{Brown97a}.

We are led to conclude from all these considerations that the adiabatic characterization of self-trapping and the self-trapping transition are strongly limited in 1-D, 2-D, and 3-D, and break down definitively in the weak-coupling regime.

This and related work suggests that the self-trapping phenomenon is qualitatively similar in 1-D, 2-D, and 3-D, and involves a smooth transition between distinct weak- and strong-coupling polaron structures having large and small radii, respectively.
The strong-coupling structure is consistent with traditional notions of small polarons, including strong coupling perturbation theory and the adiabatic approximation.
The structure of polaron states in the weak-to-intermediate coupling regime appears to differ significantly from what has traditionally been associated with the terms "large polaron" and "free", but still appears to be beyond the reach of many traditional methods.
A number of relatively new and mutually-consistent non-perturbative approaches to polaron theory afford reliable and accurate information in these areas.
We have used the Global-Local variational method to analyze 1-D polaron structure in detail throughout the polaron parameter space\cite{Brown97b,Romero98a}, and higher-dimensional calculations by such methods are in progress\cite{Romero98}.

\section*{Acknowledgement}

The authors thank Josep Porr\`{a} for useful discussions in the course of completing this work.
This work was supported in part by the U.S. Department of Energy under Grant No.
 DE-FG03-86ER13606.

\bibliography{../Bibliography/theory,../Bibliography/books,../Bibliography/experiment,../Bibliography/temporary}

\end{document}